# Dynamics of ZnO nanowires immersed in in-plane switching liquid crystal cells

Yin Tao[1,2,a)] and Yiu Ho Tam[2]

[1]Laser Physics and Nonlinear Optics, MESA+ Institute for Nanotechnology, University of Twente, 7522 NB, The Netherlands

[2]Department of Physics, The Chinese University of Hong Kong, Shatin, Hong Kong

We investigated both numerically and experimentally the dynamics of individual ZnO nanowires immersed in an in-plane switching (IPS) 4-Cyano-4'-pentylbiphenyl (5CB) liquid crystal cell under switching electric fields. Comparing the motion of nanowires captured by a high-speed CCD camera with the simulated results allows the interaction among nanowires, liquid crystals and external electric field to be studied. Our results show that in the nematic phase, the relaxation and response of a nanowire are both controlled by both the dielectrophoretic torque induced by the external electric field and the elastic torque arising from the liquid crystals.

Nanomaterials such as nanowires (NWs) have attracted tremendous interest due to their unique and fascinating properties in mechanics, electronics and optics.[1-3] However, it remains a challenge to completely manipulate them. Although a few methods have been developed such as atomic force microscopy[4] and self-assemble during growth,[5,6] the application in Nano-devices remains limited. Liquid crystals (LCs) are a collection of self-assembled uniaxial rod-shaped molecules in their nematic phase. Due to this special property, the director of LCs $\bar{n}$ can be easily controlled by a suitable alignment layer. Furthermore, it is also can be reoriented by

---

[a)]Author to whom correspondence should be addressed. Electronic mail: y.tao-1@utwente.nl

the external electric or magnetic fields. Several reports have already shown that the anisotropic elasticity of the LC introduces a longitudinal anchoring on the surface of NWs to align the Ni NWs.[7] By applying an external magnetic field, the reorientation of the NWs can be achieved.[8] Several papers have reported the ability to align, transport and rotate the Au NWs using dielectrophoretic force in deionized (DI) water.[9] Other scholars have investigated the ability of manipulation of NWs in LC by indirect observations such as measuring the photoluminescence,[10] capacitance,[11] absorbance,[12] and conductivity[13] of the LC+ NWs mixture compared with pure LC. However, these results were obtained from a comparison of the average values of above parameters in the cells before and after an external electric or magnetic field applied. Even for those direct observations mentioned above are restricted either to the magnetic material or DI water as an isotropic medium where no longitudinal anchoring exists to help restore NWs back to their original positions. Hence, we propose to find a more general method to manipulate all kinds of NWs in LCs such as aligning and reorienting. In this paper, we will demonstrate the realization of aligning and reorienting ZnO NWs immersed in 5CB by applying external electric fields. Through the study of the NW dynamics, we demonstrated the feasibility of using LCs as a medium for achieving alignment and manipulating the orientation of NWs in a controllable, repeatable, and reversible fashion with electric fields.

The LC cell we used was shown in Fig. 1. It was constructed from one finger patterned indium tin oxide (ITO) coated glass plate and one normal glass plate. These two substrates were sandwiched together with a sealant mixed with glass rod spacers (Zencatec, NEG Spacer), to form a gap thickness of about 30 $\mu$m. To align the director of LCs parallel to the electrode strips, we treated both glass plates with a thin polyimide (Zencatec model ZCMPI-410) and rubbed in the lengthwise direction parallel to the electrodes. Small amounts of ZnO NWs (1.5 wt %) fabricated by chemical vapor deposition, with lengths $L$ ranging from 5 to 20 $\mu$m, were dispersed in a positive LC 5CB (Merck. Co) in the isotropic phase at $70^oC$, and the dispersive suspensions were later introduced into a sandwich cell by capillary action. Afterwards, that cell was slowly cooled down to room temperature at an approximate rate of $2^oC/min$.

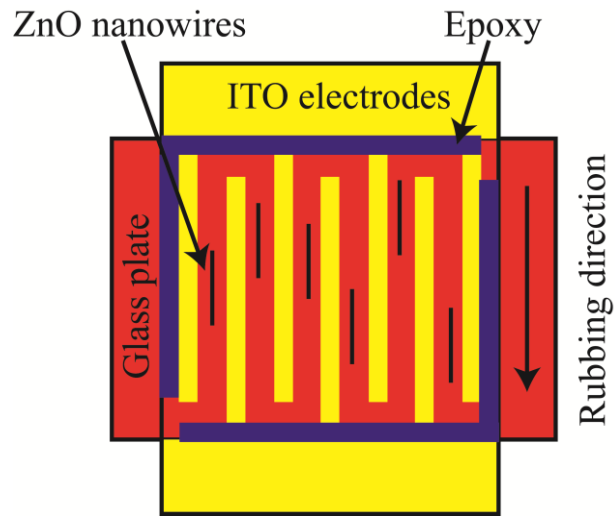

Fig. 1 Schematic of ZnO NWs immersed in an in-plane switching 5CB LC cell

All the experiments were performed at room temperature under a polarized optical microscope (Olympus BH-2) with an extra-long working distance 60X objective. A function generator combined with an AC amplifier, which can generate a maximum of 110 VAC at 1 kHz, was used to apply the external electric fields to the cell. A high-speed video camera capable of a maximum 1000 fps (FASTEC IMAGING InLine 1000) was used to record both the change of transmission in the LC cell and the dynamics of NWs.

Before we studied the dynamics of ZnO NWs immersed in LCs, the dynamics of pure 5CB LCs near the ZnO NWs was investigated. We selected the area near the ZnO NWs and far from the electrodes in the images extracted from the video (shown in Fig. 2(a) ) and measured the change of the grey scale to obtain the transmission of the LC cell during the switching in a cross polarized microscope. Fig. 2(a) shows that the NW is well aligned along the director of LCs, which is also mentioned in other papers.[7,8] and Fig. 2(b) shows the normalized transmission curve of the LC cell after removal of an external electric field of 900 $V$/mm acting on the cell.

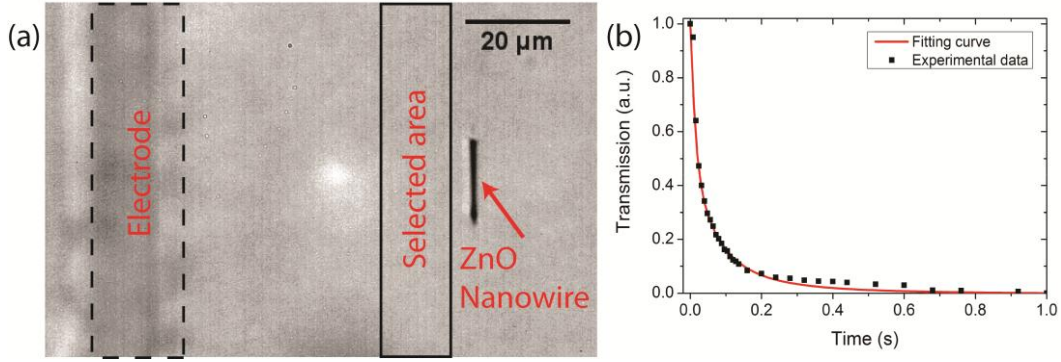

Fig. 2 (a) A typical image of a ZnO NW immersed in the cell of which the area was selected for the transmission measurement; (b) Measured and fitted normalized transmission curve of LC at the selected area after removal of the external electric field (900 $V$/mm).

The black dots in Fig 2(b) are the experimental data, while the red curve in Fig 2(b) is the fitting curve obtained by combining the solving director distribution of LCs, $\theta(z, t)$, using Leslie-Ericsson equation[14] and calculating the transmission of the cell by Jones Matrix method (details in Ref. 15).

$$\frac{\partial^2 \theta}{\partial z^2} k + \varepsilon_0 \Delta\varepsilon E^2 \sin 2\theta - \gamma_1 \frac{\partial \theta}{\partial t} = 0. \quad (1)$$

where $\theta, k, \Delta\varepsilon, \gamma_1$ are the director, elastic constant, dielectric anisotropy and rotational viscosity of the LCs, respectively.

$$T = P[\prod_m^N (S_m G_m S_m^{-1})] A I. \quad (2)$$

where $S_m G_m S_m^{-1}$ is the Jones matrix of one layer of LCs. $P$, $A$ and $I$ are the Jones matrices of the polarizer, analyzer, entrance light intensity respectively. The results of the fitting parameters ($k/\gamma_1$ and cell thickness $d$) are $3.8 \times 10^{-7}$ $dyne/P$ and 31.5 $\mu$m respectively. Compared with three reference values of 5CB[16], it was found that $k/\gamma_1$ is mostly close to the value $k_{22}/\gamma_1$ ($3.7 \times 10^{-7}$ $dyne/P$), which indicated that the dynamics of LCs in the cell is caused by the twist deformation ($k_{22}$) of the LCs.

To study the relaxation of ZnO NWs and LCs (after the removal of external electric field), we investigated the relaxation of three individual NWs of lengths (5.8 $\mu$m, 9.8 $\mu$m and 15.6 $\mu$m) after removal of various external electric fields. Fig. 3(a) shows six typical frames containing the NW3 ($L$=15.6 $\mu$m) extracted from the videos (Ref.17-19). We measured the position of the NW in terms of angle $\phi$ between the NW and rubbing direction at various times and plotted them in Fig. 3(b). The black dots indicate the relaxation of these three NWs after removal of the external electric fields of around 920 $V$/mm ($U_{rms}$=28 $V$ on the cell).

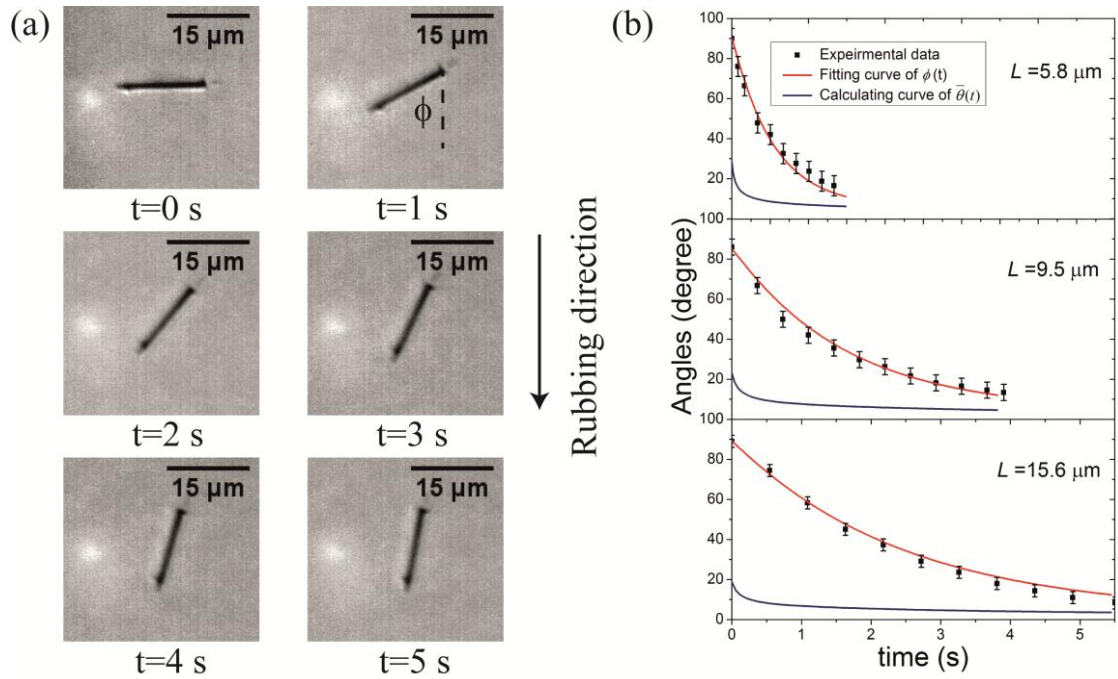

Fig. 3 (a) Six typical images extracted from the video containing the NW 3 ($L$=15.6 $\mu$m) during the relaxation after an external electric filed (920 $V$/mm) was removed. (b) Relaxation of three NWs ($L$= 5.8 $\mu$m, 9.5 $\mu$m, 15.6 $\mu$m) as a function of time. The black dots are the measurement data, and the red curves are the fitting curves of $\phi(t)$, while the blue curves are the calculating curves of the average directors of LCs around NWs $\bar{\theta}(t)$.

To quantitatively understand the physical details of the relaxation of these NWs, a simple model is proposed. The relaxation of one NW immersed in an IPS cell can be explained by following equation:

$$\frac{d\phi(t)}{dt} = \frac{4\pi Ck}{\eta}[\phi(t) - \bar{\theta}(t)]. \tag{4}$$

The term $4\pi Ck[\phi(t) - \bar{\theta}(t)]$ is the elastic torque acting on the NW. This can be derived by using an analogy between the director field of LCs and an electrostatic system.[8,20] $k$ is the elastic constant of LCs and $C$ is the equivalent electrostatic capacitance of the NW, which can be approximated by: $C \approx L/[2\ln(L/2r)]$, where $L$ and $r$ are the length and radius respectively of the NW, and $\phi(t) - \bar{\theta}(t)$ is the director difference between that of the NW and that of the far away director.[8] $\bar{\theta}(t)$ is the average twist angle of LCs around the NW. We assume the NWs are lying at the bottom of the cell due to gravity. Thus, it is mathematically defined as: $\bar{\theta}(t) = \int_0^{2r} \theta(z,t)dz/2r$, where $\theta(z, t)$ is obtained from the results of solving Leslie-Ericsson equation previously. The term $\eta d\phi(t)/dt$ is the viscous drag on the NW and the Stokes drag coefficient $\eta$, which for our NWs can be treated as an ellipsoid, is given by[21]

$$\eta = \frac{\eta_{eff}\pi L^3}{3}[\ln(\frac{L}{2r}) - 0.662] = \eta_{eff}\eta_L. \tag{5}$$

where $L$ and $r$ are the length and radius of the NW. And $\eta_{eff}$ is the effective anisotropic viscosity[22] of the LCs, which here we treated as a rotational viscosity of LC $\gamma_1$. $\eta_L$ is a purely geometrical factor for an ellipsoid in our model. Furthermore, an inertia term of the NWs is neglected, due to the small Reynolds number of the system.[23] Red curves in Fig. 3(b) are the fitting curves of $\phi(t)$ for these three individual NWs by using the model with the only fitting parameter $\chi = 4\pi Ck/\gamma_1\eta_L$, while blue curves represent the calculated values of $\bar{\theta}(t)$ for the LCs. $\phi$ decreases with time, and the decay time (the time required for $\phi$ to drop from an initial angle $\phi=90^{\circ}$ to the angle $\phi=25^{\circ}$) shown in Table 1 increases with the length of the NW. These relaxation trends of the NWs can be explained as follows: according to Eqn. 5, the viscous drag on a NW depends on $L^3$, while the elastic torque depends on $L$ through the geometrical factor $C$. Good agreement between the measured and fitted curves is found among the three NWs. The fitting parameter $\chi$ for these three NWs are 2.01 $s^{-1}$ ($L$=5.8 $\mu$m), 0.80 $s^{-1}$ ($L$=9.6 $\mu$m), 0.46 $s^{-1}$ ($L$=15.6 $\mu$m) respectively. Calculated $\bar{\theta}(t)$ is very small for all three NWs. This is because near the bottom of the cell, all the twist angles $\theta(t)$ are small as a result of the strong anchoring of LCs near the substrate. Also for the longer NW, the difference between $\phi(t)$ and $\bar{\theta}(t)$ are larger due to their smaller $\chi$.

**Table 1** Comparison of the decay times of three NWs in the IPS LC cells.

| Length of nanowires ($\mu m$) | Decay time ($s$) |
|---|---|
| 5.8 | 0.81 |
| 9.5 | 2.03 |
| 15.6 | 3.36 |

Furthermore, we also investigate the relaxation of these three NWs when $\phi(t=0)$ is smaller than 90° due to the former external electric field is not high enough to rotate the NWs to completely 90°.[24] The ratios of $k/\gamma_1$ for all these three nanowires were extracted from $\chi$ and these values are all close to $3.4 \times 10^{-7} dyne/P$. Again, comparing this with the reference value,[16] we found that $k/\gamma_1$ is close to $k_{22}/\gamma_1$. Furthermore, for all these three NWs, they always relaxed back to the direction along the rubbing direction. Therefore, we can conclude that the relaxation of NWs in the IPS LC cell is also predominantly controlled from the elastic torque induced by the twist deformation of LCs associated with the twist elastic constant $k_{22}$.

The responses of these three NWs when an external electric field varying from 100 $V$/mm to 1000 $V$/mm was applied were also studied. As are shown in Fig. 4, the black dots are the measured angles $\phi(t)$ of NW3 ($L$=15.6 $\mu$m), of which relaxation has been described previously, under three different external electric field (250, 500, 900 $V$/mm respectively), by using the same method extracted from the videos (Ref. 25-27). Comparing all the results with those for the other two NWs,[28] we found that at a higher external electric field, the NWs take less time to rotate to their equilibrium states of near 90°. Furthermore, the external electric field needs to reach more than 350 $V$/mm so that the NWs could rotate to a perpendicular orientation to the electrodes. Also, the shapes of all the curves are nearly the same, which look like an elongated '$S$' shape. It is also interesting to discover that all three NWs were always rotating in a clockwise direction, no matter what external electric field was applied. Also, the pivot of the motion was not at the center of the NWs but was located at one end. We believe that may be due to the shape and non-uniformity of mass distribution on the NW. Again, a similar model to that used previously was applied to help understand the interaction among the NWs, the LCs and the external electric field. One more term is introduced, namely a dielectrophoretic torque on the NW which caused by the external electric field, $\tau_E = V\varepsilon_m\varepsilon_0 \text{Re}[\varepsilon^*]E^2\sin 2\phi(t)/2$,[29] where $V$ is the volume of the NW, $\varepsilon_m$ is the average dielectric constant of LCs and Re[$\varepsilon$*] is the real part of the deductive dielectric constant of the ZnO NW coupled with LCs. Thus the response of the NW can be expressed as:

$$\frac{d\phi(t)}{dt} = \frac{4\pi Ck}{\eta}[\phi(t) - \bar{\theta}(t)] + \frac{1}{2\eta}V\varepsilon_m\varepsilon_0 \text{Re}[\varepsilon^*]E_{ext}^2 \sin 2\phi(t). \qquad (6)$$

By using the same fitting method as previous one except this time we set Re[$\varepsilon$*] as the only fitting parameters ($k/\gamma$ is used by the value we found during the relaxation fitting process), red fitting curves were plotted; meanwhile, $\bar{\theta}(t)$ values for the LCs were also calculated (blue curves). Again, good agreement between the measured and fitted curves is found among three NWs. Furthermore, From Eqn. 6 it can be seen that the rate of change of $\phi(t)$ is determined by both the elastic torque and dielectrophoretic torque. The dielectrophoretic torque is proportional to $E^2$ and the elastic torque depends on the average twist angle $\bar{\theta}(t)$ of the LCs, which increases as the electric fields strength increases. Hence, for higher electric fields, the elastic and dielectrophoretic torque are both larger, thus, the NWs rotate more quickly. What's more, the dielectrophoretic torque is also dependent on the factor $\sin 2\phi(t)$ that determines the shape of the curves.

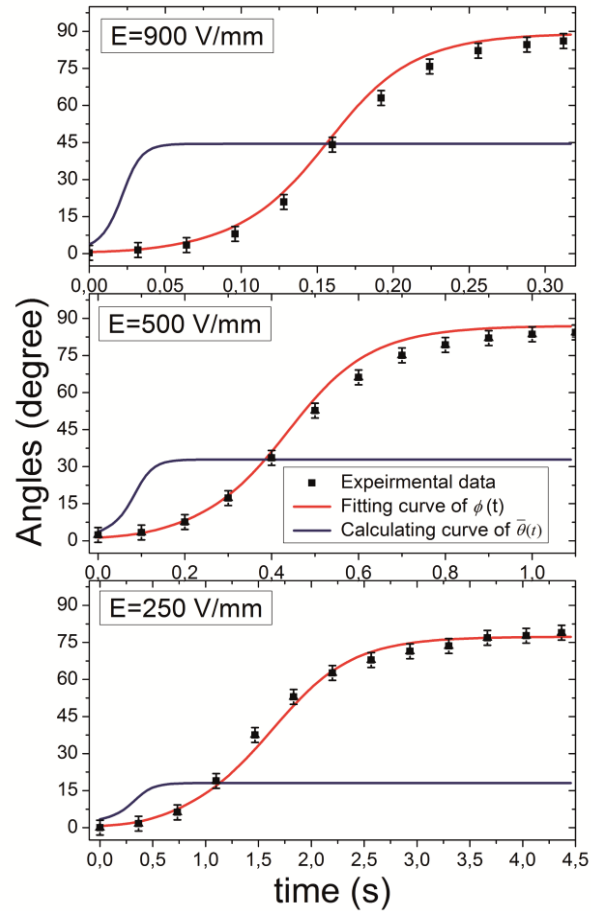

Fig. 4 Response of the NW ($L$= 15.6 $\mu$m) at three different external electric fields

($E_{ext}$=250, 500, 900 $V$/mm) as a function of time. The black dots are the measurement data, and the red curves are the fitting curve of $\phi(t)$, while the blue curves are the calculating curve of $\bar{\theta}(t)$.

The '*S*' shape of the motion of NWs can be understood from this term alone, since it dominates the motion of the NW. Also it is interesting to find that the minimum angle of the NW2 (*L*=9.5 $\mu$m) is not close to 0º and its maximum angle also does not reach 90º, even at an applied voltage as high as $E_{ext}$=430 $V$/mm. We can understand this by the fluctuations in the alignment of NWs in LCs; the same phenomenon[8] was discussed where they claimed that in the nematic phase the order parameter *S* is around 0.97, so some NWs still can't be aligned precisely by the LCs along the rubbing direction of the bottom substrate. Also this may be due to other interactions, such as the net charge on the NW and the frictional force on the NW by the bottom substrate, both of which disturb the motion of the NW. The fitting results are 13.48 (*L*=5.8 $\mu$m), 31.45 (*L*=9.6 $\mu$m), 199.52 (*L*=15.6 $\mu$m) respectively. Furthermore, we found these response and relaxation of all the NWs can be always repeated as long as the external electric field doesn't excess the breakdown field of the LCs.

In conclusion, we have investigated the dynamics of ZnO NWs immersed in an IPS LC cell. A simple model for the dynamics of ZnO NWs in LCs was proposed and the model predictions agreed well with the experimental data. For the relaxation of individual ZnO NWs, it was found that NWs follow the reorientation of the LCs, which indicates that it is the elastic torque of the latter that drives the former, which is caused by the elastic torque arising from the twist ($k_{22}$) deformation of LCs. For the response of those NWs, the interaction among ZnO NWs, LCs and the external electric field is analyzed. We confirmed that the motion of the NWs was mainly controlled by the dielectrophoretic torque acting on it rather than the elastic torque. Furthermore, the reoriented speed of NWs can be controlled by the external electric field. Thus, we confirmed that manipulation of the nanowires in the IPS LC cells can be achieved both reversely and repeatedly by applying or removing the external electric fields in nematic phase.

The authors would like to thank Prof. S.K. Hark in The Chinese University of Hong Kong for the support.